\RequirePackage{tikz}

\documentclass[pdflatex,sn-basic]{sn-jnl}

\usepackage{dsfont}
\usepackage{amsmath}

\tikzstyle{state}=[shape=circle,draw,thick,minimum size=5em]

\newcommand{\bB}{{\boldsymbol B}}
\newcommand{\bx}{{\boldsymbol x}}
\newcommand{\btheta}{{\boldsymbol\theta}}
\newcommand{\bTheta}{{\boldsymbol\Theta}}
\newcommand{\bbeta}{{\boldsymbol\beta}}
\newcommand{\bmu}{{\boldsymbol\mu}}

\jyear{2021}%

\theoremstyle{thmstyleone}%
\theoremstyle{thmstyletwo}%

\theoremstyle{thmstylethree}%

\begin{document}

\title[Modeling whale diving behavior]{Markov modeling for a satellite tag data record of whale diving behavior}

\author*[1]{\fnm{Joshua} \sur{Hewitt}}\email{joshua.hewitt@duke.edu}
\equalcont{These authors contributed equally to this work.}

\author[2,3]{\fnm{Nicola J.} \sur{Quick}}\email{nicola.quick@duke.edu}
\equalcont{These authors contributed equally to this work.}

\author[1]{\fnm{Alan E.} \sur{Gelfand}}\email{alan@duke.edu}
\equalcont{These authors contributed equally to this work.}

\author[2]{\fnm{Robert S.} \sur{Schick}}\email{robert.schick@duke.edu}
\equalcont{These authors contributed equally to this work.}

\affil*[1]{\orgdiv{Department of Statistical Science}, \orgname{Duke University}, \orgaddress{\city{Durham}, \postcode{27708}, \state{North Carolina}, \country{USA}}}

\affil[2]{\orgdiv{Nicholas School of the Environment}, \orgname{Duke University}, \orgaddress{\city{Durham}, \postcode{27708}, \state{North Carolina}, \country{USA}}}

\affil[3]{\orgname{University of Plymouth}, \orgaddress{\city{Plymouth}, \country{UK}}}

\abstract{Cuvier’s beaked whales (\textit{Ziphius cavirostris}) are the deepest diving marine mammal, consistently diving to depths exceeding 1,000m for durations longer than an hour, making them difficult animals to study.  They are important to study because they are sensitive to disturbances from naval sonar.  Satellite-linked telemetry devices provide up to 14-day long records of dive behavior.  However, the time series of depths is discretized to coarse bins due to bandwidth limitations.  We analyze telemetry data from beaked whales that were exposed to moderate levels of sonar within controlled exposure experiments (CEEs) to study behavioral responses to sound exposure.  We model the data as a hidden Markov model (HMM) over the time series of discrete depth bins, introducing partially observed movement types and recent diving activity covariates to model marginal non-stationarity.  Movement types provide more flexible modeling for CEEs than partially observed dive stages, which are more commonly used in dive behavior HMMs.  We estimate the proposed model within a hierarchical Bayesian framework, using HMM methods to compute marginalized likelihoods and posterior predictive distributions.  We assess behavioral response by comparing observed post-exposure behavior to usual unexposed behavior via the posterior predictive distribution.  The model quantifies patterns in baseline diving behavior and finds evidence that beaked whales deviate in response to sound.  We find evidence that (i) beaked whales initially shorten the time they spend between deep dives, which may have physiological effects and (ii) subsequently avoid deep dives, which can result in lost foraging opportunities.}


\maketitle

\section{Introduction}\label{sec:introduction}

Cuvier's beaked whales (\textit{Ziphius cavirostris}) are the deepest diving marine mammal, consistently making dives to depths exceeding 1,000m, for durations lasting longer than an hour \citep{tyack2006}, including some extreme records \citep{Schorr2014-md, quick2020}. The physiological limitations of this diving behaviour may make Cuvier's beaked whales particularly susceptible to disturbances from sonar; with multiple sonar events connected with strandings across the world \citep{deQuiros2019}. However, our understanding of the relationship between beaked whale diving behaviour and sonar exposure remains limited. For example, in periods of baseline diving behavior, i.e., non-exposed, Cuvier's beaked whales undertake a series of shallower, presumed non-foraging dives following a deep dive that may serve a physiological recovery purpose \citep{quick2020}. The cumulative time of these ``shallow'' dives is grouped as an inter-deep dive interval (IDDI), which encompasses the time between two deeper, presumed foraging dives \citep{tyack2006}. Beaked whales perform these dive cycles (deep foraging dive interspaced with the IDDI period) repeatedly, suggesting that underlying physiology may control or dictate the duration and extent of the cycles of dives. 

Understanding these patterns is critical to assess not only the physiological adaptations needed \citep{kooymanPhysiologicalBasisDiving1998} but also to assess the response to and the impact of acoustic disturbance on an animal \citep{Garcia_Parraga2018-zb}. For instance, the observed changes in post-exposure diving behaviour where animals  extend dive times or change surfacing patterns \citep{Tyack2011, DeRuiter2013, miller2015} presumably come with some physiological cost. Following exposure to sound, beaked whales have also been observed to lengthen the time between presumed foraging dives \citep{Falcone2017}, which may indicate flight away from the putative disturbance, and which also may carry some cost to the animal. To best estimate these costs, ideally we could first identify direct \textit{changes} to physiology such as increased heart rate and stroke rate of a diving animal and then determine if these changes are more or less likely and more or less costly depending on where the animals are in the dive cycle. However, we do not have such high-resolution data over time-scales longer than a day or two (though see \citeauthor{sweeneyCuvierBeakedWhale2022}, \citeyear{sweeneyCuvierBeakedWhale2022}), and, as such, resort to satellite telemetry devices which yield coarser, but longer, data streams. To assess if changes have occurred in these streams, we must first quantify diving cycles in the baseline diving behavior. 

We model data from satellite-linked telemetry devices that produce a discrete-time, discrete-valued time series of depths to study cycles of deep and shallow dives, and potential responses to sound exposure.  Depth data are discretized to binned values before transmission to meet satellite bandwidth constraints, and depth bins are recorded once every five minutes to maximize the duration of data collection.  Previous work with these data \citep{hewitt2021} focused on deep, presumed foraging dive behavior, isolating the deep dives within a satellite tag record, and treating them as conditionally independent in an effort to learn about baseline (absence of exposure) animal behavior during the descent, foraging, and ascent stages of a deep dive.  Here, our intent is to explore the whole sequential time series of the dives (presumed deep foraging and shallow presumed non-foraging) to better quantify changes following exposure. That is we use the entire time series of observed depths and define three diving stages: surface time stage, shallow dive stages, and deep dive stages. Using this time series of observed depths and inferred stages, we seek to learn about sequence of behaviors in the entire record.  Further, we introduce temporal covariates to learn if they can help to illuminate the sequential behavior.

In particular, our contribution is to propose a generative, nonstationary, continuous-time model for diving movement across discrete depth bins.  Nonstationarity is used to model the influence of behavioral states and is introduced through a fine-scale, discrete-time \emph{movement type} process that controls parameters for movement across depth bins. 
In different words, we offer a two-level model which provides movement types as states (i.e., fast, slow, ascent, descent)  and then random depths given the state.  Allowing states to vary at finely-spaced, discrete timepoints approximates the flexibility provided by behavior processes in continuous-time, but without posing as many computational challenges for inference \citep[cf.][]{parton2017}.

Our methodological contribution is to enrich modeling for dive data in several ways.  Customary latent-variable models have been designed to model sequences of dives observed via archival tags which record a continuous depth value once every second  \citep{langrock2013}.  By comparison, we propose a movement type model with feedback designed to model sequences of dives observed via satellite-linked telemetry tags.  Satellite-linked telemetry data differ from archival data because, due to bandwidth limitations, only discrete depth-bin values, e.g., 0--30 m,  are reported once every 5 minutes due to user defined setting choices.  We extend a latent-variable model designed to learn about characteristics of individual deep dives observed via satellite-linked telemetry devices \citep{hewitt2021}.

Our biological contribution is to gain behavioral insight from modeling 14 days of dive data all at once.  The longer time period  allows the impact of covariates to be studied across hundreds of dive cycles.  We propose constructed covariates to enable explanation of potential response.  Such explanation benefits from the longer window of study. Altogether, we are able to achieve a better understanding of baseline diving behavior over a longer study window.

Further, the approach allows for investigation of changes in diving behavior following the exposure of the animal to simulated sonar, through an experimental study known as a behavioral response study (BRS), comprised of controlled exposure experiments (CEEs) \citep{Southall2016, Harris2018-ew}. In a CEE, the experimental protocol is to 1) place a biologging device on an animal, 2) collect baseline behavior, i.e., before an exposure event, 3) expose the animal to a sound of interest, and 4) record behavior to look for a response to the exposure and ultimate return to baseline diving behavior. In this context, we introduce exposure (time of, but not strength of).  However, we do not attempt to model a change point.  Instead, we examine whether post-exposure behavior differs from unexposed/baseline behavior. 

The data reported here were collected during the Atlantic Behavioral Response Study (BRS) \citep{Southall2018-brs}, which uses high-resolution kinematic and acoustic tags \citep{JohnsonTyack2003} and Satellite tags to collect data on movements and diving of deep diving whale species exposed to sonar. Here we focus on the time series of depths recorded every 5-minutes at the individual level for approximately 14 days; detail and exploratory analysis presented in the next section (Section \ref{sec:data}). Then, we specify the two-state model for baseline diving behavior (Section \ref{sec:model}), and describe methods for posterior computation and special quantities of interest (Section \ref{sec:inference}) before presenting results (Section \ref{sec:results}).  We conclude with a discussion of the model and results, limitations, as well as directions for future work and extensions (Section \ref{sec:discussion}).

\section{The data}\label{sec:data}

The data comprise 14-day dive records of 32 individual Cuvier's beaked whales (\textit{Ziphius cavirostris}). These whales were tagged off Cape Hatteras, NC USA, as part of the Atlantic BRS \citep{Southall2018-brs}. Depth data for each individual was recorded by a SPLASH-10 tag from Wildlife Computers, Inc (Redmond, Washington, USA) (Figure \ref{fig:data_illustration}).  The depth of the animal in the water column was reported every 5 minutes for approximately 14 days due to  tradeoffs in the resolution of recording and reporting binned depth data every 5 minutes \citep{cioffi2022, Quick2019}.

We used data from 32 tags between 2018--2020 as part of the Atlantic BRS. These tags comprise data from 12 different controlled exposure experiments (CEEs), containing exposure events ranging from 7--60 minutes, where individual tagged animals were exposed to simulated or real Navy sonar.  Because the timing of each CEE was known, we intersected these with the depth time series to assign baseline, exposure, and post-exposure sections of each individual's data record. Owing to weather and field logistics, the amount of baseline dive behavior data prior to exposure varies between animals.

\section{The model}\label{sec:model}

\begin{figure}
\begin{center}
	\includegraphics[width=\textwidth]{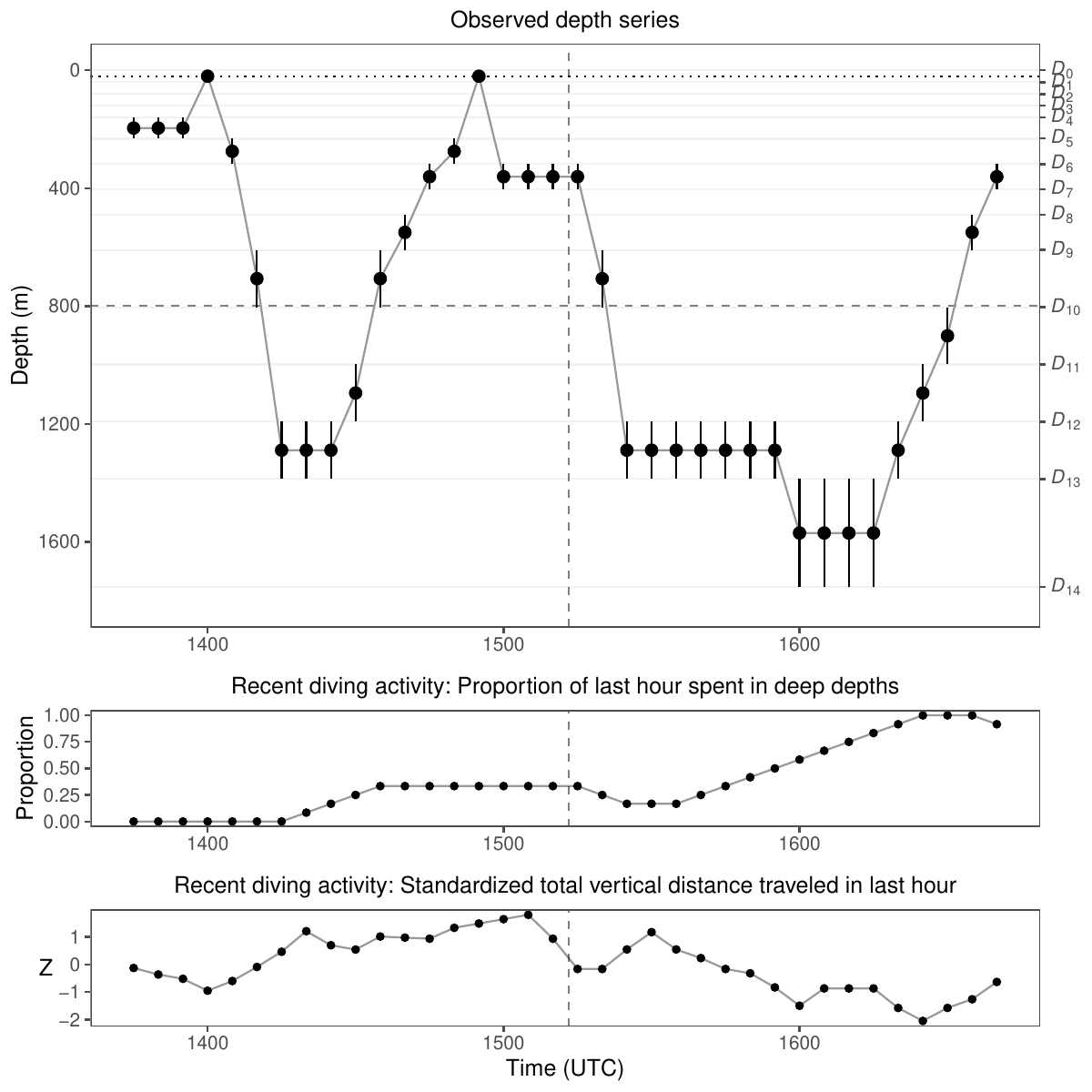}
	\caption{Dive trajectory for ZcTag087\_DUML surrounding the animal's exposure to sound at the time indicated by the vertical dotted line.  Horizontal lines denote depth bin breakpoints $D_0,\dots,D_{14}$.  Points and error bars represent the bin centers and widths associated with observations $\ell_{ij}$. Lower panels illustrate the linear terms of the recent diving activity covariates introduced in Section \ref{sec:model}.}
	\label{fig:data_illustration}
\end{center}
\end{figure}

We explicitly model the baseline portion for each tag, then use deviations from the associated posterior predictive distribution to non-parametrically assess post-exposure departures from baseline behaviors.  Each telemetry tag reports a regularly-spaced time series of binned depth observations.  Conceptually, breakpoints $D_0<D_1<\dots<D_M$ define $M$ depth bins, where $D_0=0$ is the ocean surface and $D_M$ is the maximum depth observed.  The $\ell$th depth bin $\ell\in\{1,\dots,M\}$ labels the depth range $[D_{\ell-1},D_\ell)$, whose width we denote by $d_\ell\equiv D_\ell - D_{\ell-1}$.  The tag for animal $i\in\{1,\dots,N\}$ records $T_i$ depth bin observations $\ell_{i1},\dots,\ell_{iT_i}$, with $\ell_{ij}\in\{1,\dots,M\}$ for $j\in\{1,\dots,T_i\}$ (Figure \ref{fig:data_illustration}).  The observation $\ell_{ij}$ is made a time $t_{ij}>0$, and the time between observations is constant for all $i$ and $j$, i.e.,  $\delta_{ij}=t_{i,j+1}-t_{ij} \equiv \delta$. We modify a continuous-time discrete-state model for \emph{individual} deep dives from \cite{hewitt2021} to account for \emph{all} observed dives in $\ell_{i1},\dots,\ell_{iT_i}$ and cycles between dive types (i.e., deep vs.\ shallow).

As stated, our model introduces explicit movement types to characterize transition distributions between consecutive depth bin observations, for example, from $\ell_{ij}$ to $\ell_{i,j+1}$.  The state $s_{ij}\in\{1,\dots,K\}$ is explicitly defined to represent one of $K=5$ movement types: 1) slow descent, 2) fast descent, 3) slow random walk, 4) slow ascent, and 5) fast ascent.  The explicitly-defined states differs from standard HMMs, where states are unlabeled a priori and interpreted after estimation.  Our states are explicitly defined a priori, but are unobserved (only the depth data is observed) and are estimated from the data, as latent state models for dives often do \citep{langrock2013}.  We estimate values for ``fast'' and ``slow'', but the choice of states reflects typical observations of beaked whales from high-resolution, short duration telemetry devices, in which the whales tend to ascend and descend either quickly ($\sim$1.5 m/s) or slowly ($\sim$.5--.7 m/s) \citep{tyack2006}.  A vector of covariates $\bx_{ij}\in\mathbb R^p$ influences transitions between consecutive states; for example, from $s_{i,j-1}$ to $s_{ij}$.  In particular, $\bx_{ij}$ is not observable until $t_{ij}$.

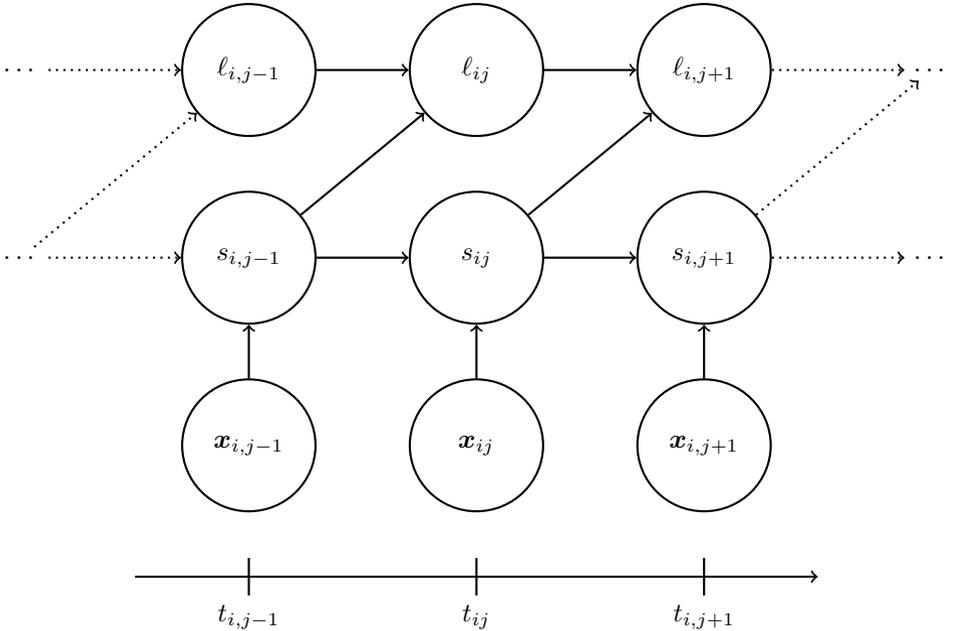
\begin{figure}
\begin{center}
\def\ybottomrow{-0.5}
\def\yrowspace{2.5}
\def\xcolzero{-1.5}
\def\xcolspace{3}
\begin{tikzpicture}[]
\draw[thick,->] (0,-2.25) -- (9,-2.25);
\foreach \x/\y in {\xcolzero+1*\xcolspace/t_{i,j-1}, 
			 \xcolzero+2*\xcolspace/t_{ij}, 
		  	 \xcolzero+3*\xcolspace/t_{i,j+1}}
    \draw[thick] (\x,-2) -- (\x, -2.5) node[below] {$\y$};
\node at(\xcolzero+4*\xcolspace,\ybottomrow+\yrowspace) (skp2) {$\dots$};
\node at(\xcolzero+4*\xcolspace,\ybottomrow+2*\yrowspace) (Lkp2) {$\dots$};
\node[state] at(\xcolzero+3*\xcolspace,\ybottomrow+\yrowspace) (skp1)  {$s_{i,j+1}$}
	edge[->,thick, dotted] (Lkp2)
	edge[->,thick, dotted] (skp2);
\node[state] at(\xcolzero+3*\xcolspace,\ybottomrow+2*\yrowspace) (Lkp1) {$\ell_{i,j+1}$}
	edge[->,thick, dotted] (Lkp2);
\node[state] at(\xcolzero+3*\xcolspace,\ybottomrow) (xp1) {$\bx_{i,j+1}$}
	edge[->,thick] (skp1);
\node[state] at(\xcolzero+2*\xcolspace,\ybottomrow+\yrowspace) (sk)  {$s_{ij}$}
	edge[->,thick] (Lkp1)
	edge[->,thick] (skp1);
\node[state] at(\xcolzero+2*\xcolspace,\ybottomrow+2*\yrowspace) (Lk) {$\ell_{ij}$}
	edge[->,thick] (Lkp1);
\node[state] at(\xcolzero+2*\xcolspace,\ybottomrow) (x) {$\bx_{ij}$}
	edge[->,thick] (sk);
\node[state] at(\xcolzero+\xcolspace,\ybottomrow+\yrowspace) (skm1)  {$s_{i,j-1}$}
	edge[->,thick] (Lk)
	edge[->,thick] (sk);
\node[state] at(\xcolzero+\xcolspace,\ybottomrow+2*\yrowspace) (Lkm1) {$\ell_{i,j-1}$}
	edge[->,thick] (Lk);
\node[state] at(\xcolzero+1*\xcolspace,\ybottomrow) (xm1) {$\bx_{i,j-1}$}
	edge[->,thick] (skm1);
\node at(\xcolzero,\ybottomrow+\yrowspace) (skm2) {$\dots$}
	edge[->,thick, dotted] (Lkm1)
	edge[->,thick, dotted] (skm1);
\node at(\xcolzero,\ybottomrow+2*\yrowspace) (Lkm2) {$\dots$}
	edge[->,thick, dotted] (Lkm1);
\end{tikzpicture}
\end{center}
\caption{Schematic of dependence between the depth bin, unobserved state, and covariate sequences, aligned with time.}
\label{fig:dependence_schematic}
\end{figure}

The use of movement types as states for modeling whale diving activity is novel compared with previous definitions of states.  Models for dive behavior typically use dive \emph{stages} to specify transitions from $\ell_{ij}$ to $\ell_{i,j+1}$ instead.  For example, such dive stages can denote the descent stage of a deep dive, the foraging stage of a deep dive, the ascent stage of a deep dive, etc.\ \citep{hewitt2021, langrock2013}.  Movement types are more local in time than dive stages because a descent movement at time $j$ does not require the longer-term context of whether the dive associated with time $j$ is a shallow dive or a deep dive.  The dependence structure for our model can be illustrated via a graphical model (Figure \ref{fig:dependence_schematic}).

The distribution for the depth bin $\ell_{i,j+1}$ is Markov with respect to $\ell_{ij}$, conditional on the state $s_{ij}$.  At $s_{ij}=k$ it is driven by a vector of parameters $\btheta_k=(\lambda^{(k)}, \pi^{(k)})\in\mathbb R^{+}\times[0,1]$ that characterizes the $k$th movement type.  We use the parsimonious, within-stage depth bin transition model and distribution from \cite{hewitt2021} (eq. 4) as our observation distribution for depth bin transitions.  The parameters $\btheta_k$ for the $k$th movement type define a continuous-time Markov chain (CTMC) over the depth bins $\{1,\dots,M\}$, in which $\lambda^{(k)}$ models the average vertical speed at which an animal moves, and $\pi^{(k)}$ models the probability that an animal moves to the next deepest depth bin when a transition occurs (i.e., from bin $\ell$ to bin $\ell+1$).  The CTMC models the average time an animal stays in bin $\ell$ before transitioning to an adjacent bin via an exponential random variable with mean $d_\ell/\lambda^{(k)}$.  So, the average dwell time in bin $\ell$ is directly proportional to the bin's width $d_\ell$ and inversely to the movement speed $\lambda^{(k)}$.    We specify the transition distribution from $\ell_{ij}$ to $\ell_{i,j+1}$ via the CTMC's finite time transition distribution, defined such that
\begin{align}
\label{eq:depth_transitions}
	P(\ell_{i,j+1}=v \vert \ell_{ij}=u, s_{ij}=k, \btheta_{k}) &= \pi_{uv}(\btheta_{k}),
\end{align}
where $\pi_{uv}(\btheta_{k})\in[0,1]$ is the $(u,v)$ entry of the matrix exponential transformation $e^{\delta A(\btheta_{k})}$ of the CTMC's transition rate matrix $A(\btheta_{k})\in\mathbb R^{M\times M}$ scaled by the time between observations $\delta$.  The distribution \eqref{eq:depth_transitions} models the net change in depth bin after an animal swims with movement type $k$ during the time interval $[t_{ij}, t_{i,j+1})$.

The distribution for the state $s_{ij}$ is Markov with respect to $s_{i,j-1}=k$, conditional on the covariates $\bx_{ij}$ and a state-dependent block-column matrix of parameter vectors that include both fixed and random effects $\bB_{ik} = [\bbeta_{ik1} \dots \bbeta_{ikK}]\in\mathbb R^{p\times K}$.  The covariates and parameters specify the transition distribution for $s_{ij}$ via a multinomial logit model \citep[Section 7.1]{agresti2002}.  Multinomial logit models specify the log-ratios of switching from state $s_{i,j-1}=k$ to $s_{ij}=l$ vs.\ not switching via
\begin{align}
\label{eq:multinomial_logit}
	\log \frac{P(
		s_{ij}=l \vert s_{i,j-1}=k, \bx_{ij}, \bB_{ik}
	)} {P(
		s_{ij}=k \vert s_{i,j-1}=k, \bx_{ij}, \bB_{ik}
	)} 
	=
	\bx_{ij}^T\bbeta_{ikl},
\end{align}
for $l=1,\dots,K$ and $\bbeta_{ikK}=\boldsymbol 0$ to ensure identifiability.  The softmax function normalizes the $K$ log-ratios to yield the transition distribution
\begin{align}
\label{eq:state_transitions}
	P(s_{ij}=l \vert s_{i,j-1}=k, \bx_{ij}, \bB_{ik}) = 
	\frac{
		\exp\{ \bx_{ij}^T\bbeta_{ikl} \}
	} {
		\sum_{k'=1}^K \exp\{ \bx_{ij}^T\bbeta_{ikk'} \}
	}.
\end{align}
The distribution \eqref{eq:state_transitions} allows the most recent diving activity and current environmental conditions, both captured via $\bx_{ij}$, to influence how the animal will move during the next time interval $[t_{ij}, t_{i,j+1})$, similar to \cite{zucchini2008}.  Across several time steps, covariates model patterns in diving behavior by influencing the relative frequency of transitions between the different movement types.

Including recent diving activity in the covariates $\bx_{ij}$ induces higher-order Markov dependence within the model.  Beaked whales tend to engage in longer periods of shallow dives following longer deep dives in an apparent need to rest and recover \citep{quick2020}.  In this regard, we introduce two choices of constructed auto-regressive covariates: (i) the proportion of observations in the past hour in depth bins whose centers are at deep depths (i.e., deeper than 800m \citep{baird2006, baird2008, Shearer2019}), and (ii) the total vertical distance traveled in the past hour (Figure \ref{fig:data_illustration}).  We sum the distances between observed depth bin centers to quantify the total vertical distance covariate.   We incorporate the recovery effect through these two predictors via separate cubic polynomials. We chose polynomials because exploratory investigation indicated that linearity was not sufficiently flexible enough for modeling state transitions.  Mathematically, let $m_\ell \equiv (D_{\ell-1} + D_\ell)/2$ denote the midpoint of depth bin $\ell\in\{1,\dots,M\}$.  With $\delta=300\text{s}$, the first covariate in $\bx_{ij}$ is a cubic polynomial based on  $c_{ij1}\equiv\sum_{k=0}^{11}\mathds{1}\{m_{\ell_{i,j-k} \geq 800}\}$; the second is a cubic polynomial based on $c_{ij2}\equiv\sum_{k=0}^{11}\vert m_{\ell_{i,j-k}} - m_{\ell_{i,j-k-1}}\vert$ after centering by 1,650m and scaling by 500m. The centering and scaling standardize each covariate to have an empirical mean and standard deviation of approximately 0 and 1, respectively.  The polynomials share a common intercept for identifiability, which we model via random effects for each individual; the remaining terms are modeled as fixed effects.  The random effects let each animal have different baseline tendencies to make transitions to and from different movement types.  They also enable borrowing of strength across the individuals.  The fixed effects assume the overall physiological processes are common between animals.

Beaked whales, in particular, are also hypothesized to use deep dives to avoid predators.  The animals are observed to spend less time near the surface during the day and on bright, moonlit nights \citep{barlow2020}.  Following \cite{barlow2020}, we include celestial effects in $\bx_{ij}$ to model the predator avoidance effect.  The vector $\bx_{ij}$ includes dummy variables that indicate whether it was daytime, a dark night, or a moonlit night at the observation time $t_{ij}$. 

Altogether, the log-ratio in \eqref{eq:multinomial_logit} becomes
\begin{align*}
    \bx_{ij}^T\bbeta_{ikl} &= \beta_{ikl0} + \\ & \phantom{=}
    \beta_{kl1} (\text{Daytime})_{ij} + 
    \beta_{kl2} (\text{Dark night})_{ij} + 
    \beta_{kl3} (\text{Moonlit night})_{ij} + \\ & \phantom{=}
    \beta_{kl4} c_{ij1} +
    \beta_{kl5} c_{ij1}^2 +
    \beta_{kl6} c_{ij1}^3 + \\ & \phantom{=}
    \beta_{kl7} z_{ij2} +
    \beta_{kl8} z_{ij2}^2 +
    \beta_{kl9} z_{ij2}^3, 
\end{align*}
for animal $i$ at time $j$ transitioning from movement type $k$ to movement type $l$, where $\beta_{ikl0}$ is the animal-specific random effect, $(\text{Daytime})_{ij}$ and related are indicator variables for celestial conditions, $z_{ij2}=(c_{ij2}-1650)/500$ is the approximately standardized version of $c_{ij2}$, and the remaining $\beta_{kl\cdot}$ terms are fixed effects.  The $\beta_{kl\cdot}$ terms do not depend on $i$ since they are common across all animals.

\section{Estimation}\label{sec:inference}

\subsection{Prior specification}

We estimate the model described in Section \ref{sec:model} within a hierarchical Bayesian framework.  Again, strength is borrowed between individuals and also between movement states via the prior specification.  The movement type parameter vectors $\btheta_1,\dots,\btheta_5$ share common elements.  We assume all ``slow'' states have common speed $\lambda_1$ such that $\lambda_1\equiv\lambda^{(1)}=\lambda^{(3)}=\lambda^{(4)}$.  We also assume all ``fast'' states have common speed $\lambda_2$ such that $\lambda_2\equiv\lambda^{(2)}=\lambda^{(5)}$.  We specify a joint prior $f(\lambda_1,\lambda_2)\propto f_\Gamma(\lambda_1;a,b)f_\Gamma(\lambda_2;a,b)\mathds{1}\{\lambda_2>\lambda_1\}$, where $f_\Gamma(\cdot,a,b)$ is the probability density function for a Gamma random variable with shape $a=.01$ and rate $b=.01$.  The joint prior orders $\lambda_1$ and $\lambda_2$ such that the ``slow'' speed $\lambda_1$ is constrained to be slower than the ``fast'' speed $\lambda_2$.  The individual Gamma densities are uninformative because they have prior means and variances of 1 and 100, respectively, which represents large prior uncertainty for vertical movement speeds of beaked whales \citep[cf.][]{tyack2006}.  We assume all ``descent'' states have common descent parameter $\pi_1$ such that  $\pi_1\equiv\pi^{(1)}=\pi^{(2)}$, and all ``ascent'' states have common ascent parameter $\pi_2$ such that $\pi_2\equiv\pi^{(4)}=\pi^{(5)}$.  We place Uniform priors on $\pi_1, \pi_2$ such that $\pi_1\sim\text{U}(0.5,1)$ and $\pi_2\sim\text{U}(0.5,1)$.  The prior distributions for $\pi_1$ and $\pi_2$ constrain the supports for the parameters to match their interpretation as ``ascent'' and ``descent'' parameters, but are otherwise uninformative.  Since we use $\pi^{(3)}$  to model random vertical movement, we assume a known $\pi^{(3)}=0.5$.

The movement type transition parameter matrices for each animal $\bB_{i1},\dots,\bB_{iK}$ have more complicated hierarchical structure.  The multinomial logit model \eqref{eq:multinomial_logit} uses the parameter vector $\bbeta_{ikl}$ to model the net effect the covariates $\bx_{ij}$ have on the relative odds that animal $i$ will switch from movement type $k$ to $l$ at time $j$.  However, the absolute probabilities \eqref{eq:state_transitions} use the full collection of parameter vectors $\bB_{ik}=[\bbeta_{ik1}\dots\bbeta_{ikK}]$ to model transitions from movement type $k$.  If an individual covariate $x_{ijr}\in\mathbb R, r\in\{1,\dots,p\}$ within $\bx_{ij}\in\mathbb R^p$ increases the relative odds of switching to one particular movement type, then it should also decrease the relative odds of switching to other movement types.  As mentioned in Section \ref{sec:model}, a subset of covariates $\mathcal S\subseteq\{1,\dots,p\}$ are associated with random effects.  We account for the relative odds relationship among each set of random effects by specifying the correlated prior distribution $[\beta_{ik1r}\dots\beta_{ikKr}]\sim\mathcal N(\bmu_{kr},\Sigma_{kr})$, where $\beta_{iklr}$ is the $r$th element of $\bbeta_{ikl}$ and $k,l\in\{1,\dots,K\}$ and $r\in\mathcal S$.  The fixed effects define $[\beta_{ik1r}\dots\beta_{ikKr}] = \bmu_{kr}$ for $r\in\{1,\dots,p\}\backslash\mathcal S$.  For all covariates $r\in\{1,\dots,p\}$, the population-level parameters have dispersed, uninformative prior distributions $\bmu_{kr}\sim\mathcal N(\boldsymbol 0, 100I)$ and $\Sigma_{kr}\sim\text{Inv-Wishart}(I/100, K)$.  

\subsection{Posterior sampling}

The likelihood for the model is specified sequentially via
\begin{align}
\label{eq:sequential_lik}
    \prod_{i=1}^N [\ell_{i1}][s_{i1}] \prod_{j=1}^{T_i-1}
        [\ell_{i,j+1}\vert\ell_{ij},\bTheta,\bB_i],
\end{align}
where $\bTheta=(\btheta_1,\dots,\btheta_K)$ and $\bB_i=(\bB_{i1},\dots,\bB_{iK})$, and $[\ell_{i1}]$ and $[s_{i1}]$ are prior distributions for animal $i$'s initial depth bin and movement type, respectively. Typically, $[\ell_{i1}]$ and $[s_{i1}]$ are specified to be discrete uniform distributions.  The sequential component in \eqref{eq:sequential_lik} uses forward-backward algorithms for Hidden Markov models (HMMs) to integrate over the state via
\begin{multline}
\label{eq:marginal_state}
    [\ell_{i,j+1}\vert\ell_{ij},\bTheta,\bB_i]
    =
    \sum_{l,k=1}^K \left(
    \pi_{\ell_{ij}\ell_{i,j+1}}(\btheta_k)
    [s_{ij}=k\vert s_{i,j-1}=l,\bx_{ij}, \bB_{il}] \times\right.\\\left.
    [s_{i,j-1}=l\vert\ell_{i,1:j-1},\bTheta,\bB_i] \right),
\end{multline}
where $\ell_{i,1:j-1}=(\ell_{i1},\dots,\ell_{i,j-1})$ and $[s_{i,j-1}=l\vert\ell_{i,1:j-1},\bTheta,\bB_i]$ is a posterior marginal distribution for the model \citep{doucet2009}.  The posterior marginal distribution in \eqref{eq:marginal_state} is evaluated recursively via
\begin{multline*}
    [s_{i,j-1}=l\vert\ell_{i,1:j-1},\bTheta,\bB_i] \propto
    \pi_{\ell_{ij-2}\ell_{i,j-1}}(\btheta_l) \times \\
    \sum_{k'=1}^K
    [s_{i,j-1}=l\vert s_{ij-2}=k',\bx_{i,j-1},\bB_{ik'}]
    [s_{ij-2}=k'\vert\ell_{1,j-2},\bTheta,\bB_i].
\end{multline*}
The hierarchical prior distribution for the model parameters is specified via
\begin{align*}
    [\lambda_1,\lambda_2][\pi_1][\pi_2]
    \prod_{k=1}^K
    \prod_{r=1}^p
        [\bmu_{kr}]
        [\Sigma_{kr}]
    \prod_{i=1}^N
        [\beta_{ik1r},\dots,\beta_{ikKr}\vert\bmu_{kr},\Sigma_{kr}].
\end{align*}

The posterior distribution is sampled via Markov chain Monte Carlo (MCMC) methods.  Conjugate updates are used when possible, in particular for the population-level centering for random effects and their covariance matrices $\{\bmu_{kr}, \Sigma_{kr}:k=1,\dots,K; r\in\mathcal S\}$.  Adaptive Metropolis-Hastings Random walk updates are used for all other parameters \citep{andrieu2008}.

\subsection{Posterior predictive simulation}

Beaked whales are hypothesized to avoid perceived threats in a similar way to their anti-predator response \citep{Tyack2011}.  One aspect of this is to initiate a deep dive, and previous CEE studies have observed beaked whales initiate deep dives in response to sonar exposure \citep{Tyack2011, DeRuiter2013, miller2015, Falcone2017, harris2017}.  However, the animals also routinely initiate deep dives, presumably to forage \citep{tyack2006}.  Disentangling whether a deep dive following exposure to sonar is a routine dive or a potential threat avoidance response motivates much of the model's design.  These are behavioral questions which cannot be answered by studying posterior distributions for model parameters directly.  

We pose two questions to use the model to help quantify potential response behaviors.  First, how do covariates influence the duration of dive cycles?  Second, does exposure to naval sonar cause animals to respond by initiating deep avoidance dives, regardless of where the animal is in their deep vs.\ shallow dive cycle?  We make these questions more precise, then use posterior predictive simulation to draw inference on the scientific quantities of interest.

For beaked whales, we define a dive cycle to be the time it takes to go from the start of one deep dive to the start of a second deep dive, similar to the length of a deep dive plus the inter-deep dive interval (IDDI) \citep{tyack2006}.  A deep dive in the area off North Carolina is any dive whose maximum depth exceeds 800m, and a shallow dive is any dive whose maximum depth does not exceed 200m \citep{baird2006, baird2008, Shearer2019}.  We study deep dive cycles via posterior distributions for threshold crossing times.  Threshold crossing times are important because we do not explicitly model dive stages \citep[cf.][]{hewitt2021, langrock2013}.

To study covariate effects, we use different combinations of covariate values to capture the time it takes simulated whales to go from the shallowest depth bin to the first bin with $m_\ell>800$. Since the covariates quantify recent diving activity from the past hour---that is, the 12 most recent observations---we use a known sequence $\ell_{01},\dots,\ell_{0,12}$ with $\ell_{0,12}=1$ to start each simulation, where the animal index $i=0$ represents a hypothetical whale.  We vary the sequence so that the full collection of simulations spans the range of values the covariate vector $\bx_{0,12}$ can attain.

Simulations are made conditionally on draws from the posterior distribution for population-level model parameters $[\bTheta,\{\bmu_{kr}\}\vert\{\ell_{ij}:i=1,\dots,N;j=1,\dots,T_i\}]$.  The simulation begins at time $j=12$ and is stopped at the first time $j=12+H_1(1)$ a deep depth has been exceeded $m_{\ell_{0,12+H_1(1)}}>800$.  We refer to $H_1(1)$ as a \emph{deep-dive hitting time} and use composition sampling to draw inference \citep[eq. 6.6]{banerjee2015}.  Specifically, we average realizations of $H_1(1)$ computed from dives simulated using the posterior samples of model parameters.  More generally, $H_1(\ell)$ for $\ell\in\{1,\dots,M\}$ is the time it takes to go from bin $\ell$ to a first depth that has exceeded 800m.  We also repeat the composition sampling process to study the posterior predictive distribution for $H_1(1)$ for each individual.  Individual-level distributions can deviate from the population-level distribution due to the random effects. 

We employ counterfactual simulation to study response behavior with respect to $H_1(\ell)$.  Such simulation uses the proposed model to provide a baseline distribution. Then, in the context of sound exposure, we use this distribution to compare the diving behavior that is observed after exposure to what could have occurred in the absence of exposure.  Simulations for animal $i$, exposed at time $j^*$, are made starting from the last pre-exposure observation $\ell_{ij^*}$ conditionally on draws from the posterior distribution for parameters, random effects, and state $[\bTheta,\bB_i,s_{ij^*}\vert\{\ell_{ij}:i=1,\dots,N;j=1,\dots,T_i\}]$.  A change in $H_1(\ell_{ij^*})$ indicates a possible, deep avoidance dive.  The hitting time $H_1(\ell_{ij^*})$ is obtained for each simulation in addition to a variation $H_2(\ell_{ij^*})$, which captures a full dive cycle.  Specifically, the time $H_2(\ell_{ij^*})$ is the time it takes to go from depth bin $\ell_{ij^*}$ to a deep depth, return to a depth shallower than 200m, and finally return to a deep depth once again.  A change in $H_2(\ell_{ij^*})$ indicates a change in regular diving behavior, which could be caused by possible horizontal avoidance, as an animal attempts to move away from the sound source \citep{DeRuiter2013,Falcone2017}.  Changes might also indicate slow ascents that have been observed post-exposure and may be potentially linked to predator avoidance behavior \citep{Tyack2011}.

We compare the pair of observed hitting times $(h_1(\ell_{ij^*}), h_2(\ell_{ij^*}))$ to the joint posterior distribution $[(H_1(\ell_{ij^*}), H_2(\ell_{ij^*}))\vert\{\ell_{ij}:i=1,\dots,N;j=1,\dots,T_i\}]$ to assess response behavior.  The model can provide evidence that animal $i$ initiated a deep dive to avoid the sound source if the posterior probability $P(H_1(\ell_{ij^*}) < h_1(\ell_{ij^*})\vert\{\ell_{ij}:i=1,\dots,N;j=1,\dots,T_i\})$ is small.  A small posterior probability indicates the observed hitting time $h_1(\ell_{ij^*})$ would be fast in the absence of sound exposure, relative to the animal's pre-exposure activity and environment.  The model can also provide evidence that animal $i$ deviated from routine deep diving behavior if, for example, the posterior probability $P(H_2(\ell_{ij^*}) - H_1(\ell_{ij^*}) < h_2(\ell_{ij^*} - h_1(\ell_{ij^*})\vert\{\ell_{ij}:i=1,\dots,N;j=1,\dots,T_i\})$ is extreme (i.e., either very large or very small).  An extreme posterior probability indicates the time between deep dives following exposure is longer or shorter than in the absence of sound exposure, relative to the animal's pre-exposure activity and environment.

\section{Results}\label{sec:results}

We ran multiple chains started from diffuse values, each until convergence was diagnosed.  Each chain is initialized with independent draws of $\pi_1$, $\pi_2$, $\lambda_1$, and $\lambda_2$ from the prior.  The parameter vectors and covariance matrices are initialized with null and identity values such that all $\bB_{ik}=\boldsymbol 0$ and $\Sigma_{kr}=I$, respectively.  The initial conditions for $\bB_{ik}$ and $\Sigma_{kr}$ initialize each chain with uniform state transition distributions \eqref{eq:state_transitions}.  We retained a posterior sample from each of the chains and pooled these samples to obtain an overall posterior sample.  Inference regarding parameters was obtained from this overall sample.  Prediction was obtained from this overall sample using composition sampling \citep[eq. 6.6]{banerjee2015}.  The posterior samples suggested multimodality for the posterior of the model parameters.  Posterior inference from this sampling is informative as we demonstrate in what follows.

\subsection{Parameter estimates}

All movement type definition parameters are estimated with high posterior precision (Table \ref{table:param_ests}). The fast speed parameter $\lambda_2$ is estimated to be faster than the slow speed parameter $\lambda_1$ by .79 m/s (95\% HPD interval: 0.76, 0.81). The posterior means for $\lambda_1$ and $\lambda_2$ are similar to the average speeds estimated for shallow and deep dives observed with higher-resolution telemetry devices, respectively \citep[Table 2]{tyack2006}. However, the higher resolution devices estimate distinct speeds for the ascent and descent stages of shallow and deep dives.

\begin{table}[ht]
\centering
\caption{Posterior distribution summaries for $\bTheta$ components.} 
\label{table:param_ests}
\begin{tabular}{cccc}
Parameter & Post. mean & Post. s.d. & 95\% HPD \\ 
  \midrule
$\lambda_1$ & 0.45 & 0.04 & (0.35, 0.51) \\ 
  $\lambda_2$ & 1.31 & 0.08 & (1.22, 1.47) \\ 
  $\pi_1$ & 0.83 & 0.09 & (0.74, 0.99) \\ 
  $\pi_2$ & 0.04 & 0.02 & (0.02, 0.08)
\end{tabular}
\end{table}

It is not possible to directly interpret posterior distributions for the stage transition covariate effects because the parameters have non-linear, interacting effects on diving behavior.  The covariate effects interact with each other due to the multinomial logit model in which the relative effects of covariates are more important than absolute effects.  Instead, we use the posterior simulations to interpret the parameters with respect to observable diving behaviors, such as the time between deep dives, which is a more scientifically meaningful quantity.

\subsection{Covariate effects}

\begin{figure}
\begin{center}
	\includegraphics[width=\textwidth]{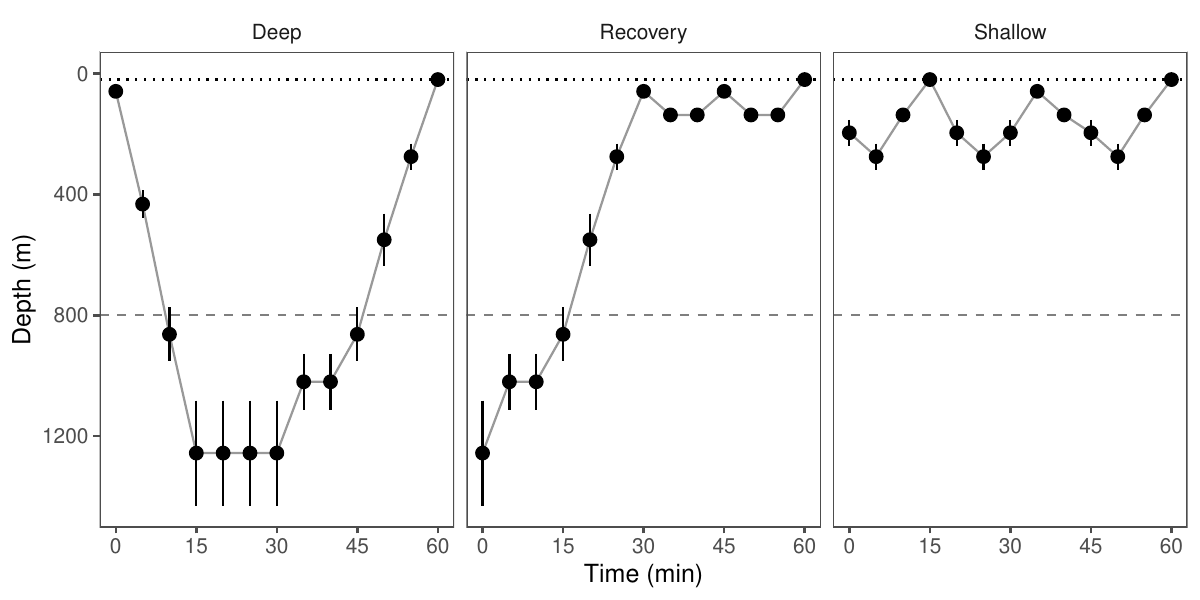}
	\caption{The three depth bin sequences $\ell_{01},\dots,\ell_{0,12}$ used to start posterior simulations for parameter interpretation, labeled with the recent diving activity they represent.}
	\label{fig:template_dives}
\end{center}
\end{figure}

We use three distinct choices for the simulation sequence $\ell_{01},\dots,\ell_{0,12}$ to explore the impact of recent diving activity covariates on the deep dive hitting time $H_1(1)$ (Figure \ref{fig:template_dives}).  These three sequences are sufficient to characterize the principal variation in recent diving activity (Supplement, Figure 3).  The first sequence is representative of animals that have just finished a deep dive, at which time the recent diving covariate values are high (Supplement, Figure 4).  The second sequence is representative of animals that have finished a deep dive at some point in the past hour and are recovering from the activity, at which time the recent diving covariate values have moderate levels.  The final sequence is representative of animals that have not been in a deep dive for more than one hour, at which time the recent diving covariate values have low levels.  We also vary the simulation starting conditions with respect to the celestial covariate and initial movement type at time $s_{0,12}$.

\begin{figure}
\begin{center}
	\includegraphics[width=\textwidth]{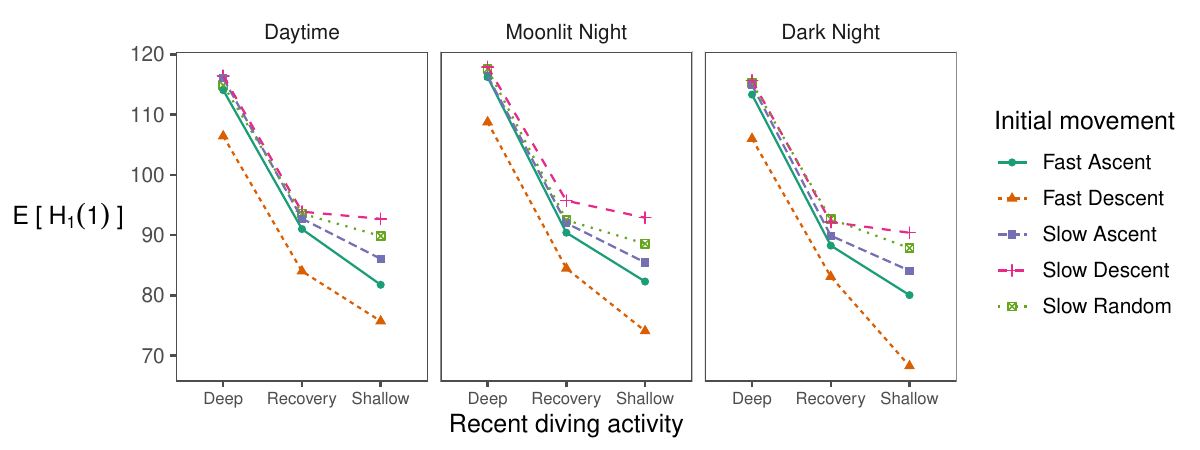}
	\caption{Posterior predictive mean for $H_1(1)$ in minutes associated with different combinations of recent diving activity, initial movement type $s_{0,12}$, and environmental conditions.}
	\label{fig:covariate_effects}
\end{center}
\end{figure}

\begin{figure}
\begin{center}
	\includegraphics[width=\textwidth]{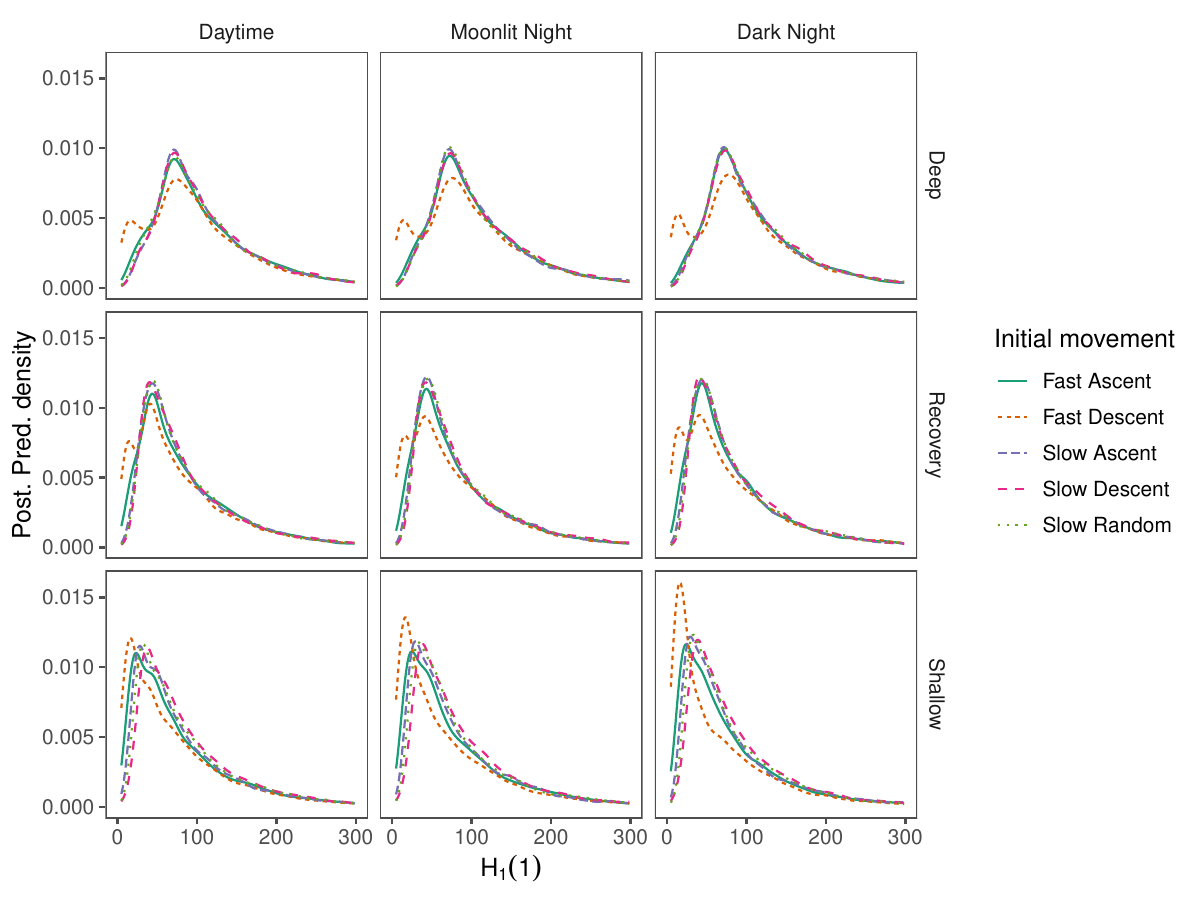}
	\caption{Posterior predictive densities for $H_1(1)$ in minutes associated with different combinations of recent diving activity, initial movement type $s_{0,12}$, and environmental conditions.}
	\label{fig:covariate_effects_var}
\end{center}
\end{figure}

Recent diving activity is the primary effect on the average value for $H_1(1)$.  Within each combination of initial movement type and celestial conditions, the model estimates that beaked whales tend to take longer on average to visit deep depths following deep diving activity than following progressively longer periods of time in shallow water (Figure \ref{fig:covariate_effects}).  Within each combination of recent diving activity and initial movement type, mean differences across celestial conditions appear to be much smaller, but variability can be larger during the day and smaller at night  (Figure \ref{fig:covariate_effects_var}).

The initial movement type $s_{0,12}$ also impacts the average value for $H_1(1)$.  Beginning simulations with fast descent movement yields the shortest average values for $H_1(1)$ because fast descent movements are most likely to be associated with the start of deep dives.  By comparison, the longest average values for $H_1(1)$ occur for simulations started with slow descent movement.  Slow descent movements are most likely associated with the start of shallow, recovery dives, so require more time to transition into the start of a deep dive.  Slow ascent and slow random walk movements have middle-range values for $H_1(1)$.  Slow ascent movements are most likely associated with the end of shallow, recovery dives, so allow more possibility to begin a deep dive.  Random walk movements are likely more ambiguously defined at shallow depths, so can also provide more possibility to begin a deep dive.  Surprisingly, fast ascent movements have relatively short $H_1(1)$ values.  The result suggests that transitions from fast-ascent/descent to fast-descent/ascent or slow-ascent/descent to slow-descent/ascent movement states tend to be more likely than transitions from fast/slow to slow/fast movement states.  The model also estimates that there is more variability in the effect of initial movement type at night than during the day if recent diving activity has been shallow.

The random effects within the stage transition model estimates individual-level variability in the location and scale of the posterior predictive distributions for $H_1(1)$.  Conditional on posterior simulation for each individual and with simulations beginning during daytime from an initial fast descent movement type, the shapes of the predictive distributions for $H_1(1)$ are similar, although their location and scale differ (Supplement, Figure 5).  The posterior predictive mean for $H_1(1)$ ranges from 55 to 128 minutes across individuals, with an average value of 90 minutes.  The posterior predictive standard deviation ranges from 52 to 102 minutes, with an average value of 77 minutes.

\subsection{Response behavior}

To assess potential deep avoidance dive responses, we focus our analysis on the 15 animals who were in shallow waters at the time of exposure.  The data are analyzed for two dive response behaviors: 1) immediate, deep avoidance dives, and 2) post-exposure avoidance of routine deep dives, which are assumed to be used for foraging.  The first response would manifest as a short time $H_1(\ell_{ij^*})$ to reach a deep depth.  The second response would manifest as a long time between deep dives, which we can measure via the difference $H_2(\ell_{ij^*})-H_1(\ell_{ij^*})$.  

While evidence for individual responses is important, evidence for population-level trends is also important.  For this data, population-level trends cannot be studied directly by estimating changepoint parameters that are added to the model.  Changepoint parameters would be poorly identified by the data because, for example, the CEEs yield only a handful of observations per animal that can be used to study immediate deep avoidance diving phenomena.  Instead, we study population-level trends through other posterior comparisons.

Tail probabilities of the posterior cumulative distribution functions (CDFs) for $H_1(\ell_{ij^*})$ and $H_2(\ell_{ij^*})-H_1(\ell_{ij^*})$ let response behaviors be compared across animals on a common, unitless scale. Transformations of the quantities $H_1(\ell_{ij^*})$ and $H_2(\ell_{ij^*})-H_1(\ell_{ij^*})$ for between-animal comparison need to account for differences between  individuals and their exposure conditions.  Although all 15 animals are exposed at a shallow depth, the exact exposure depth bin $\ell_{ij^*}$ and recent diving activity covariates $\bx_{ij^*}$ varies.  So, $H_1(\ell_{ij^*})$ and $H_2(\ell_{ij^*})-H_1(\ell_{ij^*})$ cannot be compared directly.  Tail probabilities account for all features (i.e., location, scale, skew, etc.) of the predictive distributions for $H_1(\ell_{ij^*})$ and $H_2(\ell_{ij^*})-H_1(\ell_{ij^*})$, which vary in response to exposure conditions and model parameters.  By comparison, simpler tools, such as standardized scores (i.e., z-scores) would not be adequate because the data are bounded below by 0 and higher-order moments of the CDFs vary between animals.

\begin{figure}
\begin{center}
	\includegraphics[width=\textwidth]{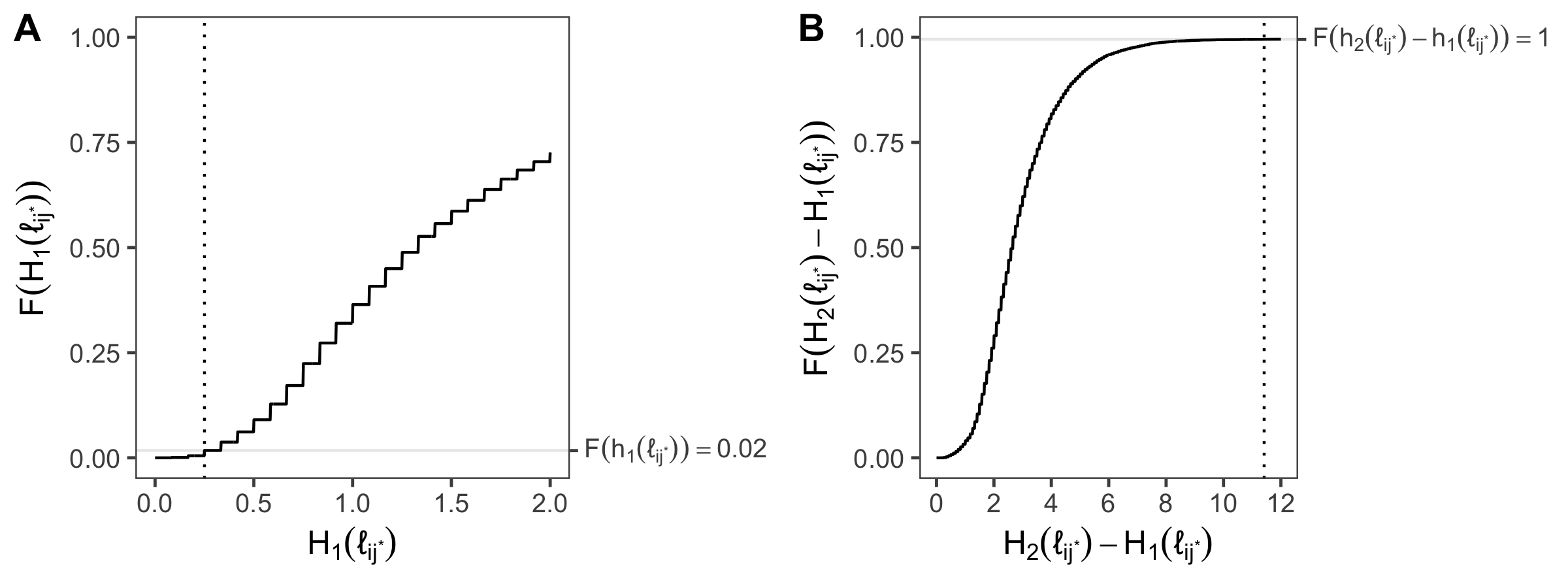}
	\caption{Posterior predictive cumulative distribution functions (CDFs) for A) first hitting time $H_1(\ell_{ij^*})$ in hours, and B) recovery period $H_2(\ell_{ij^*}) - H_1(\ell_{ij^*})$ in hours for ZcTag087\_DUML following exposure.  The vertical dotted lines denote the observed post-exposure behaviors.  The horizontal lines denote the CDF values for the observations.}
	\label{fig:zctag087_results}
\end{center}
\end{figure}

\begin{figure}
\begin{center}
	\includegraphics[width=\textwidth]{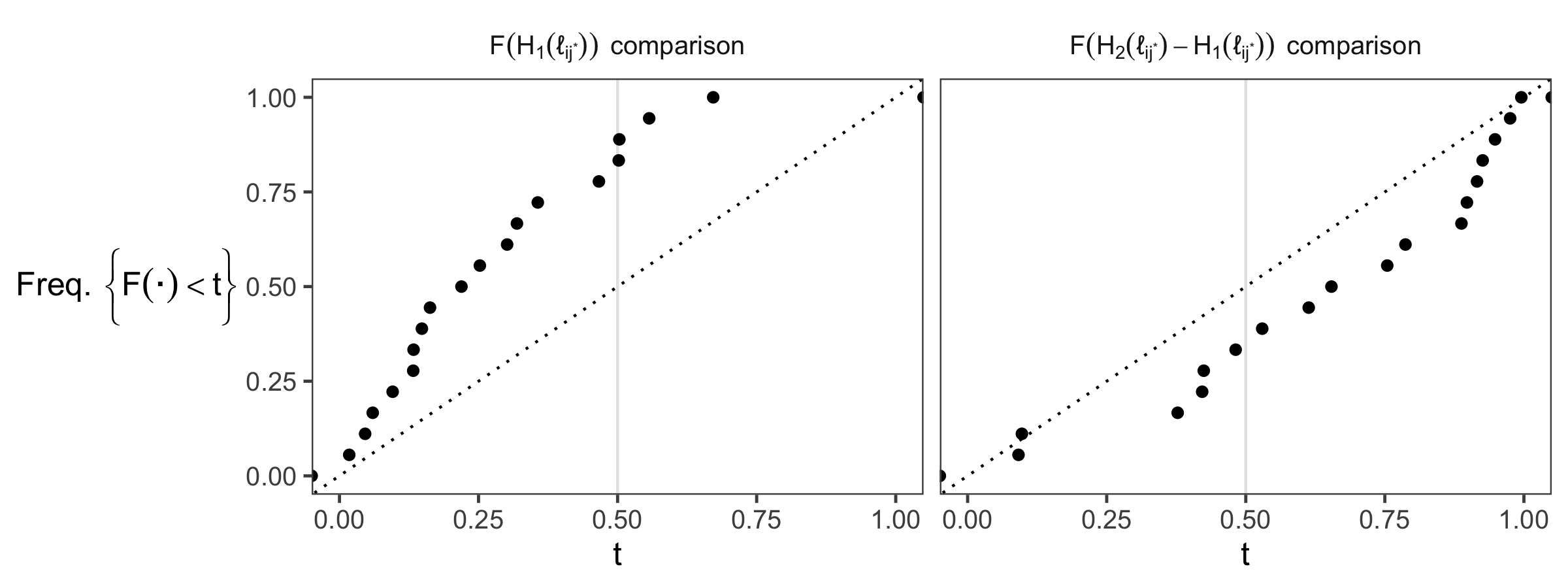}
	\caption{Empirical CDFs of posterior predictive tail probabilities for $H_1(\ell_{ij^*})$ and $H_2(\ell_{ij^*})-H_1(\ell_{ij^*})$. The x-axis shows the support for the posterior predictive tail probabilities, and the y-axis shows the frequency at which tail probabilities occur for the response behavior data.  The 1:1 reference line depicts the CDF for a Uniform(0,1) distribution.}
	\label{fig:pop_results}
\end{center}
\end{figure}

Response varies by individual, but displays an overall pattern.  The predictive CDFs for each individual are fairly diffuse, indicating that it is challenging to detect responses at an individual level (Figure \ref{fig:zctag087_results}).  A few individuals have extremely small (i.e., less than .05) tail probabilities for $H_1(\ell_{ij^*})$, and some individuals have extremely large (i.e., greater than .95) tail probabilities for $H_2(\ell_{ij^*})-H_1(\ell_{ij^*})$  (Figure \ref{fig:pop_results}).  The extreme probabilities provide evidence for individual responses that animals initiate deep, avoidance dives, and avoid routine deep diving behavior, respectively.  However, the tail probabilities for nearly all animals are underdispersed relative to an assumption that exposure to sound does not cause behavioral response.  If sound did not cause behavioral response, then the collection of tail probabilities should be uniformly distributed between 0 and 1, which is not observed (Figure \ref{fig:pop_results}).  The underdispersed tail probabilities provide evidence that exposure to sound causes animals to tend to initiate a deep dive faster than they would without exposure to sound, and avoid additional deep diving behavior for longer than they would without exposure to sound.

\section{Discussion}\label{sec:discussion}

Diving behavior is difficult to analyze from satellite-linked telemetry devices that report several weeks of discretized depth data, thus motivating model development.  Although long-duration collection captures cycles in diving behavior that are important to account for natural variation in behavior, it results in relatively limited temporal resolution due to limitations in data storage and transmission \citep{Quick2019}.  Such limited resolution makes studying immediate exposure responses during CEEs more challenging.  To reconcile these issues, we propose a model which includes features that prior work does not offer.

Modeling dive behavior with respect to movement types makes states more identifiable than models that use dive stages.   Movement types are more identifiable from data because they characterize speed and direction (i.e., ``fast ascent''), whereas dive stages characterize intent (i.e., ``ascent stage of a deep, foraging dive''). Speed and direction primarily draw on information from transitions between sequential pairs of depth bins $\ell_{ij},\ell_{i,j+1}$, whereas dive stages require information from longer sequences of depth bins for context.  

Movement types also allow a larger range of diving behavior to be modeled.  Movement types are a more fundamental description of diving behavior than dive stages, so can be combined to capture novel diving patterns, such as behavioral responses during CEEs.  Deep dive responses during CEEs imply that animals may make \textit{within-dive} transitions from shallow to deep dives.  Stage-based models for diving behavior typically do not allow within-dive transitions between dive types.  That is, such models do not allow transitions from a shallow-dive descent to a deep-dive descent, which could be typical of a behavioral response.  Modeling dive behavior with respect to movement types makes such transitions possible.  For example, transitioning from a shallow-dive descent to a deep-dive descent manifests as a transition from slow descent movement to fast descent movement, which is supported in the movement-type model we propose.  Conditional on recent diving activity, such a transition may be unlikely, which helps to identify the transition as a behavioral response to sound exposure.

Several species of whales--the most prominent of which are the speciose family of Beaked whales--have been observed to change their diving and swimming behavior following exposure to sound. Many of our modeling choices are guided by previous studies in which beaked whales have changed their behavior following such exposure; these include \citet{Tyack2011, DeRuiter2013, miller2015, Falcone2017,isojunnoWhenNoiseGoes2020a, wensveenNorthernBottlenoseWhales2019}. Many of these studies have used short-term, higher resolution biologging devices to record behavior, or if they use longer-term devices, they are typically programmed to record only dive duration and maximum observed depth (this type of satellite tag setting is known as the ``Behavior'' setting on some tags \citep{Quick2019}). In contrast, our analysis uses long-duration time series of depth bin data (i.e., several weeks) to refine quantification of potential behavioral responses by accounting for cycles in diving behavior.  Cycles of diving behavior are harder to identify from short duration time series (i.e., several days) or long duration dive summary data, such as the data streams from Behavior tags. As such, we have attempted to better quantify our understanding of these responses and of what comprises baseline behavior. 

Our model quantifies covariate effects on dive cycles and finds evidence for behavioral response during CEEs.  The posterior predictive distribution for dives finds that the mean time between deep dives is primarily controlled by recent diving activity.  Celestial conditions appear to have a smaller contribution, perhaps mainly influencing variability in the time between deep dives.  Individual responses to sound exposure are difficult to identify from 5-minute observations of discretized depth data because the (counterfactual) predictive distribution for non-exposed behavior is relatively diffuse for most animals.  However, the data from multiple individuals provide evidence for a larger trend that post-exposure diving behavior is not consistent with predictions of non-exposed behavior.

The covariates capturing recent diving activity allow dive cycles to be modeled, but also make estimation challenging.  A multinomial logit model for transitions between movement types uses the recent diving activity covariates.  Multinomial logit models require a large number of parameters, which makes estimation computationally expensive.  The large parameter space also makes convergence of MCMC samplers difficult.  We addressed the latter issue by combining output from multiple chains which converge to approximately the same parameter values.

Our model can potentially be extended to analyze dive data from other species.  The model parameters are challenging to estimate and the model defines a relatively limited set of movement types.  The latter point is a modeling choice made for the beaked whales we study.  Beaked whales have fairly regular diving behavior, which does not require a large number of movement types to describe.  Other species may require defining more speed and direction categories, which are straightforward modifications of the model.  Similarly, other species may require additional covariates for recent diving behavior.  The focus of our model on estimating baseline behavior also allows a wide range of potential behavioral responses to be studied.  As with our application, arbitrary response behaviors can be studied by comparing observed post-exposure behaviors to the distribution of expected behaviors using posterior predictive simulation.

\backmatter

\bmhead{Supplementary information}

Supplementary information with additional details about the MCMC sampler and posterior simulations are available.

\bmhead{Acknowledgments}

We thank William R.\ Cioffi for stimulating conversation and assistance with data interpretation.  We thank the following for stimulating conversation including Richard Glennie, Catriona Harris, Th\'eo Michelot, Len Thomas, and Andreas Fahlman.  Computing was performed on the Duke Compute Cluster at Duke University. We thank Andy Read and Brandon Southall for use of the data.

\section*{Statements and Declarations}

\subsection*{Funding}

The research reported here was financially supported by the United States Office of Naval Research grant N000141812807, under the project Phase II Multi-study Ocean acoustics Human effects Analysis (Double MOCHA).  The Atlantic BRS provided data collection and is supported by the U.S. Navy's Marine Species Monitoring Program under Contract No.\ N62470-15-D-8006, Task Order 18F4036, Issued to HDR, Inc. 

\subsection*{Competing Interests}

The authors do not have financial or non-financial interests that are directly or indirectly related to the work submitted for publication.

\subsection*{Ethics approval, data, and consent to participate}

The data analyzed here, and all research activities carried out, were collected as part of the Atlantic Behavioral Response study under NOAA/NMFS Scientific Research Permit No.\ 14809 issued to Douglas P. Nowacek, and NOAA General Authorization letter of confirmation No.\ 16185 issued to Andrew J. Read in accordance with the relevant guidelines and regulations on the ethical use of animals as experimental subjects. The research approach was approved by the Institutional Animal Use and Care Committees (IACUC) of Duke University.

\subsection*{Code availability}

Code and data required to reproduce the analysis are made available as supplementary material at \url{https://github.com/jmhewitt/dsdive_fulltag}.

\subsection*{Authors contributions}

JH wrote and ran software to analyze the data, and contributed to the design and writing of this work.  NQ, AG, and RS contributed to the design and writing of this work.  All authors read and approved the final manuscript.

\bibliography{hewitt}

\end{document}